\documentclass[a4paper,11pt]{article}
\usepackage{jcappub} 
\usepackage{lineno}
\usepackage{bm}
\usepackage{amsmath}
\usepackage[dvipsnames]{xcolor}
\usepackage{comment}
\usepackage[normalem]{ulem}
\DeclareUnicodeCharacter{2212}{\ensuremath{-}}
\DeclareUnicodeCharacter{3B5}{$\epsilon$}

\newcommand{\thet}{{\bm{\theta}}}

\newcommand{\omeg}{{\bm{\omega}}}
\newcommand{\xx}{{\bm{x}}}
\newcommand{\bs}{{\bm{S}}}
\newcommand{\ie}{{\textsl{i.e.}}~}

\newcommand{\appref}[1]{\hyperref[#1]{App.~\ref*{#1}}}

\newcommand{\Swyft}{\texttt{Swyft}}
\newcommand{\darksirens}{\texttt{darksirens}}
\newcommand{\Sireeni}{\texttt{Sireeni}}

\definecolor{plum}{rgb}{0.56, 0.27, 0.52}
\definecolor{forestgreen}{rgb}{0.13, 0.55, 0.13}

\title{Cosmology in the Einstein Telescope era: comparing traditional and simulation-based methods for population inference}

\author[a,b]{Giovanni Antinozzi,}
\author[c]{Guillermo Franco Abell\'an,}
\author[d,e]{Davide Sciotti,}
\author[d,e]{Matteo Martinelli}
\affiliation[a]{SISSA, via Bonomea 265, 34136 Trieste, Italy}
\affiliation[b]{INFN - Sezione di Trieste, Via Valerio 2, 34127 Trieste, Italy}
\affiliation[c]{Instituto de F\'isica Corpuscular (IFIC), CSIC-Universitat de València,
Parc Cient\'ific UV, c/ Catedr\'atico Jos\'e Beltr\'an, 2, E-46980 Paterna (València), Spain}
\affiliation[d]{INAF - Osservatorio Astronomico di Roma, via Frascati 33, 00040 Monteporzio Catone
(Roma), Italy}
\affiliation[e]{INFN - Sezione di Roma, Piazzale Aldo Moro, 2 - c/o Dipartimento di Fisica, Edificio G.
Marconi, I-00185 Roma, Italy}

\emailAdd{giovanni.antinozzi@sissa.it}
\emailAdd{g.francoabellan@ific.uv.es}
\emailAdd{davide.sciotti@inaf.it}
\emailAdd{matteo.martinelli@inaf.it}

\abstract{ The next generation of gravitational wave detectors, such as the Einstein Telescope (ET), will observe orders of magnitude more binary black hole mergers than current facilities. Most of these events will lack an electromagnetic counterpart, also known as dark siren events, yet will still enable percent-level cosmological constraints. However, the likelihood traditionally used in Hierarchical Bayesian Inference (HBI) for population-level analyses becomes computationally prohibitive as the size of dark siren catalogues and population parameters grow. In this work we compare HBI against simulation-based inference (SBI) as a scalable alternative for cosmological population inference in the ET era. Studying a proof-of-concept example of a mock ET inference, we build a catalogue of $O(10^4)$ binary black hole events, then perform inference on the Hubble constant $H_0$ and matter density $\Omega_m$ in a flat $\Lambda$CDM cosmology, using both a hierarchical analytical likelihood and Marginal Neural Ratio Estimation (MNRE).
We find excellent agreement between the two approaches, with SBI reproducing the HBI posteriors to high accuracy, while requiring orders of magnitude less computation once the simulation and training cost is amortized. We further demonstrate that SBI extends straightforwardly to a joint cosmology-plus-astrophysics analysis, simultaneously constraining $(H_0,\Omega_m)$ together with the parameters of the star formation rate density, at negligible additional cost compared to the significant increase in complexity such an extension would require within the HBI framework. Our results indicate that SBI is a promising and scalable tool for population inference with third-generation GW detectors.

\vspace*{10pt} \noindent \textbf{\texttt{GitHub}}:\@ The code \Sireeni~used to perform the SBI analyses is publicly available \href{https://github.com/GFAbellan/Sireeni.git}{here}. \textbf{\texttt{GitLab}}: The code implementing the corresponding HBI analyses is publicly available \href{https://gitlab.com/gantinoz/darksirens_hbi}{here}.

}

\begin{document}
\maketitle
\flushbottom

\section{Introduction}\label{sec:intro}
Since the first observed Gravitational Wave (GW) event by the LIGO collaboration in 2015 \cite{LIGOScientific:2016aoc}, a new observational window has opened onto the Universe. From this initial detection, the number of GW events has continuously increased, with the latest Gravitational Wave Transient Catalog-5 (GWTC-5) from the LIGO-Virgo-KAGRA collaboration providing 390 confirmed detections \cite{LIGOScientific:2026jgl, LIGOScientific:2026uyd, LIGOScientific:2026ctl, LIGOScientific:2026sit, LIGOScientific:2026wfs}. The availability of such a growing number of GW events has allowed the study of astrophysical phenomena \cite{KAGRA:2021vkt, KAGRA:2021duu, LIGOScientific:2020ibl, LIGOScientific:2020ufj, LIGOScientific:2025rid, LIGOScientific:2020stg, LIGOScientific:2025brd}, the nature and properties of compact objects \cite{LIGOScientific:2016vlm, LIGOScientific:2016sjg, LIGOScientific:2017bnn, LIGOScientific:2017ycc, LIGOScientific:2017vwq, LIGOScientific:2017vox, LIGOScientific:2020aai, LIGOScientific:2020stg, LIGOScientific:2020zkf, LIGOScientific:2020iuh, LIGOScientific:2021qlt, LIGOScientific:2025rsn,LIGOScientific:2025rid, LIGOScientific:2025brd}, as well as allowing tests of General Relativity (GR) in extreme environments \cite{LIGOScientific:2016lio, LIGOScientific:2018dkp, LIGOScientific:2020tif, LIGOScientific:2021sio, LIGOScientific:2026fcf, LIGOScientific:2026qni, LIGOScientific:2026wpt, LIGOScientific:2019fpa, LIGOScientific:2025wao}.

In addition to this, GW observations have been exploited in order to address the open problems of the cosmological model. In particular, the use of GWs to constrain the cosmological model has allowed to measure the current expansion rate of the Universe, the Hubble constant $H_0$, independently of current electromagnetic constraints \cite{LIGOScientific:2017adf, LIGOScientific:2025jau, LIGOScientific:2021aug, LIGOScientific:2026uyd}. Given the tension between the early-universe \cite{Planck:2018vyg} and the local distance‑ladder determination of $H_0$ \cite{Riess:2021jrx}, which has now reached the $5-6\sigma$ level \cite{CosmoVerseNetwork:2025alb}, it has become crucial to adopt new and independent approaches to measure the Hubble constant, with GWs offering a fundamental tool to improve our understanding of the source of such a discrepancy.

Indeed, GW observations provide one crucial quantity needed to test the cosmological model, the luminosity distance $d_L$. This quantity is tightly related to the expansion history of the Universe and, if supported by a measurement of the redshift $z$, allows to constrain the cosmological model as well as measure the current expansion rate. Although information on the luminosity distance can be extracted directly from an observed GW event, the same cannot hold for the redshift, as the latter is completely degenerate with a particular combination of the progenitor system's component masses, known as the \emph{chirp mass} $\mathcal{M}$ \cite{Maggiore:2018sht}.

This is a fundamental problem for the use of GWs in Cosmology. The redshift information can be easily extracted only in the case in which a GW is observed with an electromagnetic counterpart. In this case, one refers to the GW events as \emph{bright sirens}. At the present time, only one of such events has been observed \cite{LIGOScientific:2017zic}, but it was shown how even with such a reduced statistics one is able to constrain $H_0$ \cite{LIGOScientific:2017adf} and also obtain information on extended cosmological models \cite{Ezquiaga:2017ekz,Creminelli:2017sry}. If no electromagnetic counterpart is observed, one has to rely on different methods to obtain redshift information, which are called \emph{dark sirens}. Common dark sirens methods are further divided into: \emph{spectral sirens} \cite{Taylor:2011fs,Farr:2019twy,Mastrogiovanni:2021wsd}, where information on population distributions such as the mass spectrum or the redshift distribution of the events are used to \emph{effectively} measure the redshifts; \emph{cross-correlations methods}, where one performs a statistical association of GW events with observed matter proxies, like galaxies \cite{Pierra:2025fgr} or neutral hydrogen \cite{Dupletsa:2026uqs}, whose redshift and matter distribution can be mapped; and finally, \emph{Love sirens}, which exploit single-event measurements of tidal deformabilities from binary neutron stars or black hole neutron star systems to break the mass-redshift degeneracy with the help of assumptions on the equation of state for nuclear matter \cite{DelPozzo:2015bna}.

Dark siren methods are being continuously refined, driven by the growing number of GW events available to test and improve them. However, this shift from single-event to population-level analysis has introduced its own challenges. The most common statistical tool to understand population properties is given by \emph{Hierarchical Bayesian Inference} (HBI) \cite{Mandel:2018mve, Vitale:2020aaz}, which traces back the hierarchy from measured GW parameters up to the \emph{population parameters} that regulate the underlying population distribution of true parameters, e.g. the redshift distribution of binary black hole events. The evaluation of the hierarchical likelihood, typical of HBI, is generally computationally demanding and prone to statistical and systematic biases \cite{Talbot:2023pex,Gair:2022zsa}, and several efforts are being made to make this method scalable for a vast parameter space and for large catalogues characteristic of future GW surveys \cite{Tagliazucchi:2026dpr}.

To overcome this issue, we rely in this work on Simulation Based Inference (SBI). In recent years, SBI \cite{Cranmer:2019eaq} has established itself as a powerful framework for performing Bayesian inference in challenging cosmological settings (see e.g.~\cite{SimBIG:2023ywd,DES:2024xij,vonWietersheim-Kramsta:2024cks, Novaes:2024dyh, Zeghal:2024kic, Nguyen:2024yth, Zubeldia:2025qlt, FrancoAbellan:2024tbj,Savchenko:2025jzs}). Rather than requiring an explicit analytic likelihood, SBI learns the joint distribution between parameters and data directly from forward simulations, in which the likelihood is only defined implicitly. This shift in perspective brings several advantages. First, SBI naturally enables amortized inference, making rapid re-analysis and cross-validation straightforward. It also allows systematic effects to be included at the level of the simulator, even when such effects would be prohibitively expensive to compute or intractable to write down in a closed-form likelihood. Moreover, nuisance parameters can be marginalized over simply by sampling them in the forward model, so that marginal posteriors for the parameters of interest can be obtained efficiently.

Existing SBI approaches to GW population inference treat cosmology as one component of a full high-dimensional population model, and learn per-event representations, either from single-event summaries \cite{Leyde:2023iof} or directly from GW strain data via transformers \cite{Leyde:2026hvm}. Using a simplified setup, our work is the first to perform SBI on $O(10^4)$ merger events expected from an Einstein Telescope mock catalogue, benchmarking it directly against the classical hierarchical likelihood of HBI on a cosmology-focused problem. Beyond this, our work differs from previous studies in two aspects. First, on the modelling side, we build redshift distribution models from a convolved star formation rate density instead of a single phenomenological model for redshift dependence, where the former method can be found in population-synthesis approaches for constructing redshift distributions of binary black hole mergers \cite{Santoliquido:2022kyu}. Second, on the SBI side, we employ Marginal Neural Ratio Estimation (MNRE) \cite{Cranmer:2015bka,Miller:2021hys} instead of the Neural Posterior Estimation (NPE) \cite{papamakarios_npe, Jeffrey:2020itg} used in past works. While NPE requires an estimation of the normalized probability density \cite{Alsing:2018eau, Alsing:2019xrx}, MNRE turns inference into a simple binary classification task, thus allowing more flexibility in its network architecture.

Our paper is organised as follows. In \autoref{sec:data_ET} we describe the synthetic data used to test our pipelines, together with the theoretical model for the population of GW events that we aim to constrain. \autoref{sec:analytical} and \autoref{sec:SBI} detail the two approaches under comparison, respectively the hierarchical Bayesian likelihood and the SBI method based on MNRE. In \autoref{sec:results} we compare the HBI and SBI posteriors on the cosmological parameters $(H_0,\Omega_m)$ for the case of fixed astrophysical parameters, and further present the SBI results obtained when the astrophysical parameters are simultaneously inferred. Finally, in \autoref{sec:conclusions} we summarize our main findings and discuss future prospects.

\section{Gravitational wave events and future detectors}\label{sec:data_ET}

The goal of this work is to compare the traditional statistical approach of \emph{hierarchical Bayesian inference} with the more recent \emph{simulation-based inference} approach. The constraining power of HBI generally improves with the number of observations, but so does its computational cost, making it a particularly demanding method for the large catalogues expected from the next generation of interferometers.

At present, the planned third generation gravitational-wave detectors consist of four proposed projects that will be operating from the second half of the 2030s to beyond the 2040s and capable of spanning GW signal frequencies from MHz to tens of $\mu$Hz: the \emph{Laser Interferometer Space Antenna} (LISA), the first space-based laser interferometer sensitive to the lowest frequencies \cite{LISA:2024hlh}; the \emph{Lunar Gravitational Wave Antenna} (LGWA), the first detector of planetary length with intermediate frequency sensitivity \cite{Ajith:2024mie}; and finally two earth-based detectors for the intermediate to high frequencies, \emph{Cosmic Explorer}, the longest L-shaped interferometer ever built with a 40 km arm \cite{Evans:2021gyd}; and the \emph{Einstein Telescope} (ET), the European initiative for the next generation \cite{ET:2019dnz}. In this work we focus on the ET and its reference design as three nested interferometers of 10 km arms' length in a equilateral triangle configuration.

\begin{figure}[h]
    \centering
    \includegraphics[width=0.9\linewidth]{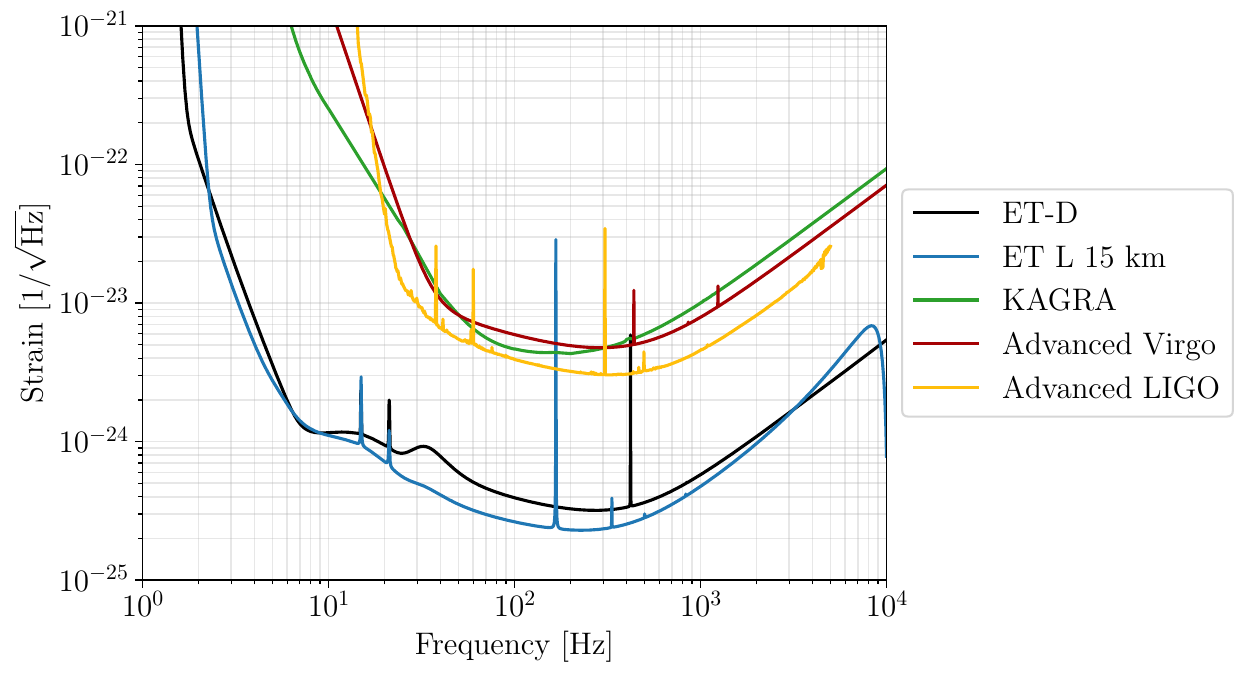}
    \caption{Amplitude spectral density of current generation interferometers compared to the Einstein Telescope. The latter is shown in two different designs: i) the arm of one of the three nested interferometers in the 10 km arms triangle (ET-D reference design), ii) one of the two widely separated Ls with 15 km arms' length.}
    \label{fig:ASD_curves}
\end{figure}

In \autoref{fig:ASD_curves} we present the projected sensitivity curve for a single detector of the triangular ET configuration in its latest ET-D model; in particular, the detector is divided into two interferometers: a cryogenic low frequency instrument (ET-LF) and a room temperature high frequency one (ET-HF) \cite{ET:2019dnz}. For comparison in figure~\ref{fig:ASD_curves} we also show the current sensitivities of the LIGO-Virgo-KAGRA collaboration interferometer and additionally we plot the alternative 15 km arm L-shaped geometry for the ET. The latter geometry for a single or two widely separated interferometers of 10,15 and 20 km arms' length has been systematically compared to the triangle geometry by Ref. \cite{Branchesi:2023mws} and it is clear that the most pronounced advantage manifests in the luminosity distance estimation, where two Ls of 15 km with a relative arm misalignment of $45$ degrees achieves better accuracy than a single-location triangular interferometer. Therefore, in the context of this paper the choice of a triangular geometry sets our baseline results from which any improvements on the luminosity distance measurements with a different ET geometry can represent an improvement in our final estimates.

\subsection{Modelling of GW events}\label{sec:thfeats}

In order to obtain a simulated dataset for ET, we first model the properties of the astrophysical sources it is expected to observe. These modelled events are then used both to simulate observational data and to generate the simulations required for the SBI approach.

A first, fundamental ingredient is the redshift distribution of the binary systems that can produce observable GW events, which is directly linked to the luminosity distance distribution we are going to exploit for our analysis. We follow the approach of \cite{Martinelli:2022elq,Antinozzi:2023yvl}, focusing only on systems composed by two black holes of stellar origin (BH), for which we assume a monochromatic mass distribution, with $M_{\rm BH}=7\,M_\odot$. It is important to stress that while this assumption on the mass should not be extremely significant for the redshift distribution modelling, a more realistic mass distribution for the BHs will affect the error calculations and the probability of detection of single events.

The redshift distribution, in the case of an ideal detector, can be obtained as \cite{Antinozzi:2023yvl}:
\begin{equation}\label{eq:zprob}
    P(z) = \frac{T_{\rm obs}R(z)}{N_{\rm id}}\,,
\end{equation}
where $T_{\rm obs}$ is the total time of observation, $R(z)$ is the merger rate of the BH systems, and $N_{\rm id}$ is the number of events that would be detected by an infinitely sensitive instrument during an observation time $T_{\rm obs}$, i.e.
\begin{equation}
    N_{\rm id} = T_{\rm obs}\int_0^\infty{{\rm d}z'\,R(z')}\,.
    \label{eq:N_id}
\end{equation}
Therefore, the probability distribution of the GW events reduces to the normalized merger rate
\begin{equation}
    P(z) = \frac{R(z)}{\int_0^\infty{{\rm d}z'\,R(z')}}\,.
\end{equation}

The merger rate can be modelled by combining the expected physical properties of the systems with the fact that they are embedded in an expanding Universe. It can be shown that the merger rate depends on the merger rate density $\mathcal{R}(z)$ and the comoving volume $V_c(z)$ as \cite{LIGOScientific:2016ebi,LIGOScientific:2017zid}
\begin{equation}\label{eq:merger_rate}
    R(z) = \frac{\mathcal{R}(z)}{1+z}\frac{{\rm d}\,V_c}{{\rm d}\,z}\,
\end{equation}
where ${\rm d}\,V_c/{\rm d}\,z$ represents the comoving volume of a spherical shell between $z$ and $z+{\rm d}\,z$ around the detector and the $(1+z)^{-1}$ term is used to convert between source frame and the detector frame. The term related to the comoving volume can be expressed in term of fundamental cosmological quantities
\begin{equation}
    \frac{{\rm d}\,V_c}{{\rm d}\,z} = \dfrac{d_L^2(z)}{(1+z)^2}\frac{4\pi c}{H(z)}\,,
\end{equation}
with $c$ the speed of light, $H(z)$ the expansion rate of the Universe and $d_L(z)$ the luminosity distance, which, for a flat Universe, can be obtained as
\begin{equation}
    d_L(z) = (1+z)\int_0^z{{\rm d}\,z'\frac{c}{H(z')}}\,.
\end{equation}

The merger rate density encapsulates our knowledge of stellar evolution which affects the life span of stars and their probability to evolve towards a BH in a binary system. Following \cite{Martinelli:2022elq}, we assume $\mathcal{R}(z)$ to be proportional to the Star Formation Rate (SFR) density $\psi_{\rm SFR}$, which we model as \cite{Nagamine:2003bd}
\begin{equation}
    \psi_{\rm SFR} = \nu\frac{a\,\exp{\left[b(z-z_m)\right]}}{a-b+b\exp{\left[a(z-z_m)\right]}}\,,
    \label{eq:SFR}
\end{equation}
where $\nu$, $a$, $b$, and $z_m$ are free parameters of the model that can be fitted by observations.

It is therefore clear how \autoref{eq:merger_rate}  contains a dependence on both SFR and cosmological parameters, which can in principle be inferred from observational data using HBI (see \autoref{sec:analytical}) or the SBI approach (see \autoref{sec:SBI}). 
 
Finally, every merger event has, in addition to its mass and redshift, additional features that do not contain information of interest for this work. These are its position in the sky, its polarisation, and the inclination of the orbital plane, the former two entering as modulation terms in the gravitational wave amplitude through the detector antenna patterns and the latter as a geometric projection factor that can directly affect the error on the luminosity distance \cite{Maggiore:2018sht}. For each simulated event, we draw the values of these parameters from uniform distributions.

\subsection{Observational error and noisy simulations}\label{sec:datasim}

Together with the features described above, simulating a GW observation also requires modelling the uncertainty with which they are measured. Since we are only interested to the luminosity distance distribution of the events, we will follow here an approximated approach, avoiding the calculation of the uncertainty for all the event's parameters. We approximate the uncertainty on the luminosity distance of an event $i$ as \cite{Cai:2016sby}
\begin{equation}\label{eq:errdl}
    \sigma_{i} = 2\frac{d_L(z_i)}{\rho_i}\,,
\end{equation}
where $\rho_i$ is the optimal signal-to-noise ratio (SNR) of the event.

Overall, we perform this calculation using the public code \texttt{darksirens}\footnote{\href{https://gitlab.com/matmartinelli/darksirens}{\texttt{https://gitlab.com/matmartinelli/darksirens}}} \cite{Martinelli:2022elq}, which implements the SNR and uncertainty calculation, as well as the theoretical modelling described in \autoref{sec:thfeats}.

In practice, a dataset is generated following the steps below:
\begin{enumerate}
    \item The redshift $z_i$ of an event is extracted from the probability distribution $P(z)$ of \autoref{eq:zprob}.
    \item Given a set of cosmological parameters, this value is converted to the ``true'' luminosity distance $d_L(z_i)$ using the public code \texttt{CAMB}\footnote{\href{https://camb.readthedocs.io/}{\texttt{https://camb.readthedocs.io/}}} \cite{Lewis:1999bs,Howlett:2012mh}.
    \item The position on the sky, the wave polarisation, and the system inclination are extracted from uniform distributions, while the mass of each binary component is set to $M_{\rm BH}$.
    \item The code computes the SNR associated to the event given ET specifications using \texttt{PyCBC} \cite{alex_nitz_2024_10473621PyCBC} and the \texttt{IMRPhenomD} waveform model \cite{Husa:2016IMRPhenomD1, Khan:2016IMRPhenomD2} and, given a threshold for this value ($\rho_{\rm thr} = 12$ in the following), it decides if the event can be observed or not (details on this calculation can be found in App. B of \cite{Martinelli:2022elq}).
    \item In the former case, the uncertainty $\sigma_i$ on the luminosity distance is computed following \autoref{eq:errdl}.
    \item The ``observed'' luminosity distance $d_L^i$ is then drawn from a normal distribution with mean $d_L(z_i)$ and variance $\sigma_i^2$.
\end{enumerate}

Repeating this procedure for a given number of input events $N_{\rm id}$, one would end up with a dataset composed of $N_{\rm obs}$ observations after the SNR cut, each with its observed luminosity distance $d_L^i$ and uncertainty $\sigma_i$.

In the following, this whole pipeline is used both to generate our simulated observed dataset and to generate the simulations required for the SBI approach. For the former, in \autoref{tab:fiducial} we show our choice for the fiducial parameters\footnote{In addition to the parameters presented in \autoref{tab:fiducial}, \texttt{darksirens} also requires to set parameters used to model primordial black holes (PBH). Here, we set the fraction of PBH to be $f_{\rm PBH}=10^{-9}$, which makes the impact of such events completely negligible.}$^{,}$\footnote{The SFR normalization parameter $\nu$ does not appear in \autoref{tab:fiducial} as one of the parameters since we set $\mathcal{R}_{\rm birth}(z)\propto \psi(z)$ (see Ref. \cite{Antinozzi:2023yvl}) and we normalize the local merger rate density to the LVK observation as explained in \cite{Martinelli:2022elq}}.

\begin{table}[h!]
    \centering
    \begin{tabular}{||c|c||c|c|c|c||}
    \hline
        \multicolumn{2}{||c||}{Cosmology} &  \multicolumn{4}{c||}{BH parameters}\\
    \hline
    $\Omega_m$ & $H_0$ [km s$^{-1}$ Mpc$^{-1}$] & $M_{\rm BH}$ [$M_\odot$] & $z_m$ & $a$ & $b$ \\
    \hline
    0.32 & 67 & 7 & 2  & 2.37 & 1.8\\
    \hline
    \end{tabular}
    \caption{Set of fiducial parameters used to generate the simulated dataset for ET.}
    \label{tab:fiducial}
\end{table}

\section{The traditional approach: hierarchical Bayesian likelihood}\label{sec:analytical}

In this section, we present a concise derivation of the hierarchical Bayesian likelihood, following \cite{Mandel:2018mve} and \cite{Vitale:2020aaz}. In an ideal scenario the measured data can directly inform on the true binary parameters ${\omeg_i}$, (i.e. distance, masses, spins, inclination, sky position, coalescence time, polarization angle)  of each of the observed events, $i = 1,...,N_{\rm obs}$. From these parameters we can infer the \emph{population parameters} $\thet$ (i.e. cosmological and astrophysical parameters) using the Bayes' theorem and the likelihood
\begin{equation}
    p(\omeg_1, ..., \omeg_{N_{\rm obs}}|\thet) = \prod_{i=1}^{N_{\rm obs}} \dfrac{p(\omeg_i|\thet)}{\int \text{d}\omeg\,p(\omeg|\thet)}\,,
    \label{eq:theta_like}
\end{equation}
where we consider independent events and even though the probabilities are normalized, i.e. $\int \text{d}\omeg\,p(\omeg|\thet)$ = 1, we explicitly write these integrals for what follows.

In the absence of noise, there might still be a detection threshold and this implies that selection effects have to be considered also in the case of an ideal detector, where ideal just means that we can measure the true binary parameters and there is no noisy scatter in their measurement. Selection effects in this case will act as a binary process, events can either be detected ($p({\rm det})=1$) or not ($p({\rm det})= 0$) depending on the binary parameter, thus we define a detection function $p({\rm det}|\omeg)$. Including the selection effects amounts to transforming the probabilities $p(\omeg|\thet)$ into $p({\rm det}|\omeg)p(\omeg|\thet)$, i.e. the probability that we measure the binary with parameters $\omeg$ and it is detectable; hence we have
\begin{equation}
    p(\{\omeg_i\}|\thet) = \prod_{i=1}^{N_{\rm obs}} \dfrac{p({\rm det}|\omeg_i)\,p(\omeg_i|\thet)}{\int \text{d}\omeg\,p({\rm det}|\omeg)\,p(\omeg|\thet)} = \prod_{i=1}^{N_{\rm obs}} \dfrac{\,p(\omeg_i|\thet)}{\int \text{d}\omeg\,p({\rm det}|\omeg)\,p(\omeg|\thet)}\,,
\end{equation}
where we used $p({\rm det}|\omeg_i) = 1$ true by construction on detected events.

In a general context, our data $\xx$ will contain both true parameters and detector noise, and a detection statistic on the data $\rho_{\xx}$ (the optimal SNR in our case)\footnote{Other additional metrics like the False Alarm Rate are also used to determine whether a signal is a true event \cite{Zheng2021Needle}.} is used to decide whether it contains a signal or not. We partition the entire dataset $\bm{\mathcal{X}}$ in two non overlapping parts depending on a threshold on $\rho_{\xx}$:
\begin{equation}
    \bm{\mathcal{X}}_{\uparrow} \equiv \rho_{\xx} > \rho_{\rm thr}\text{ and }\bm{\mathcal{X}}_{\downarrow} \equiv \rho_{\xx} < \rho_{\rm thr}\,.
\end{equation}
We can therefore partition $p({\rm det}|\omeg)$ into two dataset contributions
\begin{equation}
    p({\rm det}|\omeg) = \int_{\bm{\mathcal{X}}} \text{d}\xx\,p({\rm det}|\xx)\,p(\xx|\omeg) = \int_{\bm{\mathcal{X}}_{\uparrow}} \text{d}\xx\,p({\rm det}|\xx)\,p(\xx|\omeg) + \int_{\bm{\mathcal{X}}_{\downarrow}} \text{d}\xx\,p({\rm det}|\xx)\,p(\xx|\omeg)\,,
\end{equation}
and by construction $p({\rm det}|\xx)=1\,(0)$ if $\xx\in\bm{\mathcal{X}}_{\uparrow}$ ($\bm{\mathcal{X}}_{\downarrow}$), thus
\begin{equation}
    p({\rm det}|\omeg) = \int_{\xx\in \text{detectable}} \text{d}\xx\,p(\xx|\omeg)\,.
    \label{eq:p_det}
\end{equation}
We can now rewrite \autoref{eq:theta_like} for noisy data: 
\begin{equation}
    p(\xx_1, ..., \xx_{N_{\rm obs}}|\thet) = \prod_{i=1}^{N_{\rm obs}} \dfrac{p(\xx_i|\thet)}{\int \text{d}\xx\,p(\xx|\thet)}\,,
\end{equation}
then apply a decomposition of $p(\xx_i|\thet)$ in the true parameters
\begin{equation}
   = \prod_{i=1}^{N_{\rm obs}} \dfrac{\int \text{d}\omeg\,p(\xx_i|\omeg)\,p(\omeg|\thet)}{\int\text{d}\xx\,p(\xx|\thet)} = \prod_{i=1}^{N_{\rm obs}} \dfrac{\int \text{d}\omeg\,p(\xx_i|\omeg)\,p(\omeg|\thet)}{\int_{\bm{\mathcal{X}}_{\uparrow}}\text{d}\xx\,p(\xx|\thet)}\,,
   \label{eq:like_x}
\end{equation}
and in the last expression we use the fact that the normalization is only computed on detectable events, being $\xx_i$ one outcome in this set.

We can rewrite \autoref{eq:like_x} decomposing again with true parameters the term $p(\xx|\thet)$ in the normalization and then make use of \autoref{eq:p_det}:
\begin{equation}
    p(\{\xx_i\}|\thet) = \prod_{i=1}^{N_{\rm obs}}\dfrac{\int \text{d}\omeg\,p(\xx_i|\omeg)\,p(\omeg|\thet)}{\int_{\bm{\mathcal{X}}_{\uparrow}}\text{d}\xx\left[\int \text{d}\omeg\,p(\xx|\omeg)\,p(\omeg|\thet)\right]} = \prod_{i=1}^{N_{\rm obs}}\dfrac{\int \text{d}\omeg\,p(\xx_i|\omeg)\,p(\omeg|\thet)}{\int \text{d}\omeg\,p({\rm det}|\omeg)\,p(\omeg|\thet)}\,.
    \label{eq:shape_like_terms}
\end{equation}

Furthermore, we have to take into account the random occurrence of the events, i.e. the probability that exactly $N_{\rm obs}$ events were observed, this comes with a Poisson distribution
\begin{equation}
    p(N_{\rm obs}|N_{\rm det}) \propto e^{-N_{\rm det}} N_{\rm det}^{N_{\rm obs}}\,,
    \label{eq:number_like}
\end{equation}
where $N_{\rm det}$ is the number of \emph{events that can be detected} according to a chosen model (and its population parameters $\thet$) and the properties of a detector, and it can be found by incorporating selection effects into \autoref{eq:N_id}:
\begin{equation}
    N_{\rm det} = T_{\rm obs}\int{{\rm d}\omeg\,p(\text{det}|\omeg)R(\omeg|\thet)}\,,
\end{equation}
where $R(\omeg|\thet)$ is a general expression for the merger rate density.

Finally, we combine the Poisson contribution with the terms from \autoref{eq:shape_like_terms} and we find the \emph{hierarchical Bayesian likelihood} for noisy data:
\begin{equation}
    \mathcal{L}(\xx_1, ..., \xx_{N_{\rm obs}},N_{\rm obs}|\thet)\propto e^{-N_{\rm det}} N_{\rm det}^{N_{\rm obs}} \prod_{i=1}^{N_{\rm obs}}\dfrac{\int \text{d}\omeg\,p(\xx_i|\omeg)\,p(\omeg|\thet)}{\int \text{d}\omeg\,p({\rm det}|\omeg)\,p(\omeg|\thet)}\,.
    \label{eq:hier_like}
\end{equation}

\subsection{Simplifying the likelihood}
In the following we specialize \autoref{eq:hier_like} to the simplified case in which the only unknown binary parameter is $d_L$ and the others are fixed as described in \autoref{sec:datasim}.

It should be noted that fixing the source-frame masses in the construction of the mock dataset requires some care at the inference stage: since the source-frame chirp mass is then known, inverting the detector-frame chirp mass, directly measured from the GW waveform, could in principle allow the events' redshifts to be inferred directly, bypassing the statistical inference the method is designed to perform. We stress that this information is never propagated into our inference stage: the source-frame masses are used here exclusively to compute a more realistic SNR, and hence a more realistic luminosity-distance error, for each mock event, while the redshift inference itself is performed using only the detector-frame chirp mass and luminosity distance, exactly as it would be in a real analysis. The fixed-mass assumption therefore affects only the realism of the simulated measurement uncertainties, not the validity of the inference.\footnote{This simplification is primarily a concern for the HBI approach, which additionally requires the mass spectrum to be carefully included in the inference in order to avoid several biases. SBI is comparatively less exposed to it, since a full mass distribution can be incorporated directly into the forward simulator without a large increase in complexity, although one would still need to check for potential model misspecification in the assumed mass distribution.} Relaxing this assumption would require drawing the source-frame masses from a mass spectrum and a mass ratio distribution, increasing the realism but also the computational cost of the simulated luminosity-distance errors, rather than changing the inference method itself.

With these simplifications, the model for the binary parameters is given by
\begin{equation}
    p(\omeg|\thet) = p(d_L|\thet)\delta(\tilde{\omeg} - \tilde{\omeg}_0)\,,
\end{equation}
where we defined $\tilde{\omeg}$ as the true parameters except luminosity distance and $\tilde{\omeg}_0$ their fixed values; therefore, we can integrate \autoref{eq:hier_like} over $\tilde{\omega}$ and obtain
\begin{equation} 
    \mathcal{L}(d_L^{(1)}, ..., d_L^{(N_{\rm obs})},N_{\rm obs}|\thet)\propto e^{-N_{\rm det}} N_{\rm det}^{N_{\rm obs}} \prod_{i=1}^{N_{\rm obs}}\dfrac{\int \text{d}d_L\,p(d_L^{i}|d_L)\,p(d_L|\thet)}{\int \text{d}d_L\,p({\rm det}|d_L)\,p(d_L|\thet)}\,,
    \label{eq:hier_like_dL}
\end{equation}
and integration constants are absorbed into the normalization.

For a direct comparison with our redshift model from \autoref{sec:thfeats} we use the change of variable $d_L\rightarrow z$ and the fact that $\text{d}d_L\,p(d_L|\thet) = \text{d}z\,P(z|\thet) $ in \autoref{eq:hier_like_dL}:
\begin{equation}
    \mathcal{L}(d_L^{(1)}, ..., d_L^{(N_{\rm obs})},N_{\rm obs}|\thet)\propto e^{-N_{\rm det}} N_{\rm det}^{N_{\rm obs}} \prod_{i=1}^{N_{\rm obs}}\dfrac{\int \text{d}z\,p(d_L^{i}|d_L(z|\thet))\,P(z|\thet)}{\int \text{d}z\,p({\rm det}|z,\thet)\,P(z|\thet)}\,,
    \label{eq:hier_like_z}
\end{equation}
where $p({\rm det}|z,\thet)$ is a short-hand notation for $p({\rm det}|d_L(z|\thet))$. If we express $N_{\rm det}$ in terms of $R(z|\thet) = N_{\rm id}\,P(z|\thet)$, we find
\begin{equation}
    N_{\rm det} = \int \text{d}z\,p(\text{det}|z,\thet)\,R(z|\thet) = N_{\rm id}  \int \text{d}z\,p(\text{det}|z,\thet)\,P(z|\thet) = N_{\rm id}\,\xi(\thet)\,,
\end{equation}
where we have introduced the fraction of events above the detection threshold
\begin{equation}
    \xi(\thet) = \dfrac{N_{\rm det}}{N_{\rm id}}.
\end{equation}
We can therefore rewrite \autoref{eq:hier_like_z} in terms of its logarithm as\footnote{We do not simplify the Poissonian term with the rest of the likelihood in order to be able to show the contribution from number counts to the entire hierarchical posterior.}:
\begin{equation*}
    \log\mathcal{L}(d_L^{(1)}, ..., d_L^{(N_{\rm obs})},N_{\rm obs}|\thet) = -N_{\rm det}+N_{\rm obs}\log N_{\rm det}+\sum_{i=1}^{N_{\rm obs}}\log\left(\int \text{d}z\,p(d_L^{i}|d_L(z|\thet))\,P(z|\thet)\right)
\end{equation*}
\begin{equation}
    - N_{\rm obs}\log\xi(\thet)\,.
    \label{eq:final_hier_like}
\end{equation}
\subsection{Fraction of detectable events}\label{sec:xi}
One of the key ingredients in \autoref{eq:final_hier_like} is $\xi(\thet)$, which contains the selection effects due to the specific detector considered. Lacking an analytical expression for the detection function, we obtain $\xi(\thet)$ through an \emph{injection campaign} and a Monte Carlo approximation.

First of all, we rewrite the integral expression for $\xi$ appearing in \autoref{eq:hier_like_dL} using the Bayes' theorem as follows:
\begin{equation}
    \xi(\thet) = \int \text{d}d_L\,p(\text{det}|d_L)\,p(d_L|\thet) = \int \text{d}d_L\,\dfrac{p(d_L|\text{det}) p(\text{det})}{p(d_L)}\,p(d_L|\thet)\,,
    \label{eq:csi_def}
\end{equation}
then by finding $N_{\rm thr}$ \emph{detectable} events and approximating $p(\text{det})\approx N_{\rm thr}/N_{\rm gen}$ given that there are $N_{\rm thr}$ detectable events out of a total of $N_{\rm gen}$ generated events, we obtain from the Monte Carlo approximation:
\begin{equation}
    \xi(\thet) \approx \dfrac{1}{N_{\rm gen}}\sum_{j=1}^{N_{\rm thr}} \dfrac{p(d_{L,j}|\thet)}{p(d_{L,j})} = \dfrac{1}{N_{\rm gen}}\sum_{j=1}^{N_{\rm thr}} \dfrac{P(z_j|\thet)}{p(z_j|\thet)}\,,
    \label{eq:csi_mc}
\end{equation}
where in the last step we switched to our distribution model.

The distribution from which the events are extracted, $p(d_L|\text{det})$ or $p(z|\text{det},\thet)$, is arbitrary and the most efficient choice from a computational point of view is exactly $P(z|\thet)$. After the sampling, one retains only the events that pass the detection threshold, such that the samples effectively come from a $p(z|\text{det},\thet)$ and respect \autoref{eq:csi_mc}, this plus the fact that $p(z|\thet)=P(z|\thet)$ simplifies the expression of $\xi(\thet)$ to
\begin{equation}
    \xi(\thet) = \frac{N_{\rm thr}}{N_{\rm gen}}\,,
\end{equation}
which is again the fraction of detectable events, but specifically sampled from $P(z|\thet)$.

Generating the large number of injections required at each $\thet$ point is in general computationally demanding. We therefore adopt an importance-sampling and resampling approach, starting from injections generated at a single reference parameter $\tilde{\thet}$, we reweight and resample them to obtain injections for a generic parameter $\thet$:
\begin{itemize}
    \item[1)] Extract $N_{\rm gen}$ $z_j$ redshifts from $P(z|\tilde{\thet})$ and compute the SNRs $\rho_j = \rho(d_L(z_j|\tilde{\thet}))$.
    
    \item[2)] Compute the weights $w_j = P(z_j|\thet) / P(z_j|\tilde{\thet})$ and normalize them $\bar{w}_j = w_j / \sum_j w_j$.
    
    \item[3)] Resample $N_{\rm gen}$ new redshifts $\bar{z}_j$, drawing the $j$-th sample with probability $\bar{w}_j$; the resulting set is then distributed as an extraction from $P(z|\thet)$. Using the same weights, and exploiting the one-to-one mapping $z_j\leftrightarrow\rho_j$, we resample the $\rho_j$ into the corresponding SNRs $\bar{\rho}_j = \rho(d_L(\bar{z}_j|\tilde{\thet}))$.
    
    \item[4)] Since the SNR scales with distance as $\rho \propto d_L^{-1}$ \cite{Antinozzi:2023yvl}, we can rescale $\bar{\rho}_j$ to the correct parameters via $$\rho(d_L(\bar{z}_j|\thet)) = \bar{\rho}_j \times d_L(\bar{z}_j|\tilde{\thet}) / d_L(\bar{z}_j|\thet)\,,$$ and then determine $N_{\rm thr}$ simply by counting how many of these rescaled SNRs lie above threshold.
\end{itemize}
It is worth noting that the inclusion of additional parameters (e.g masses) would imply an higher dimensional importance sampling (point 2) using a model $p(\omeg|\thet)$ and a more general SNR scaling relation (point 4), both adding a larger computational cost which is outside the scope of this paper.

\begin{figure}[h]
    \centering
    \includegraphics[width=1\linewidth]{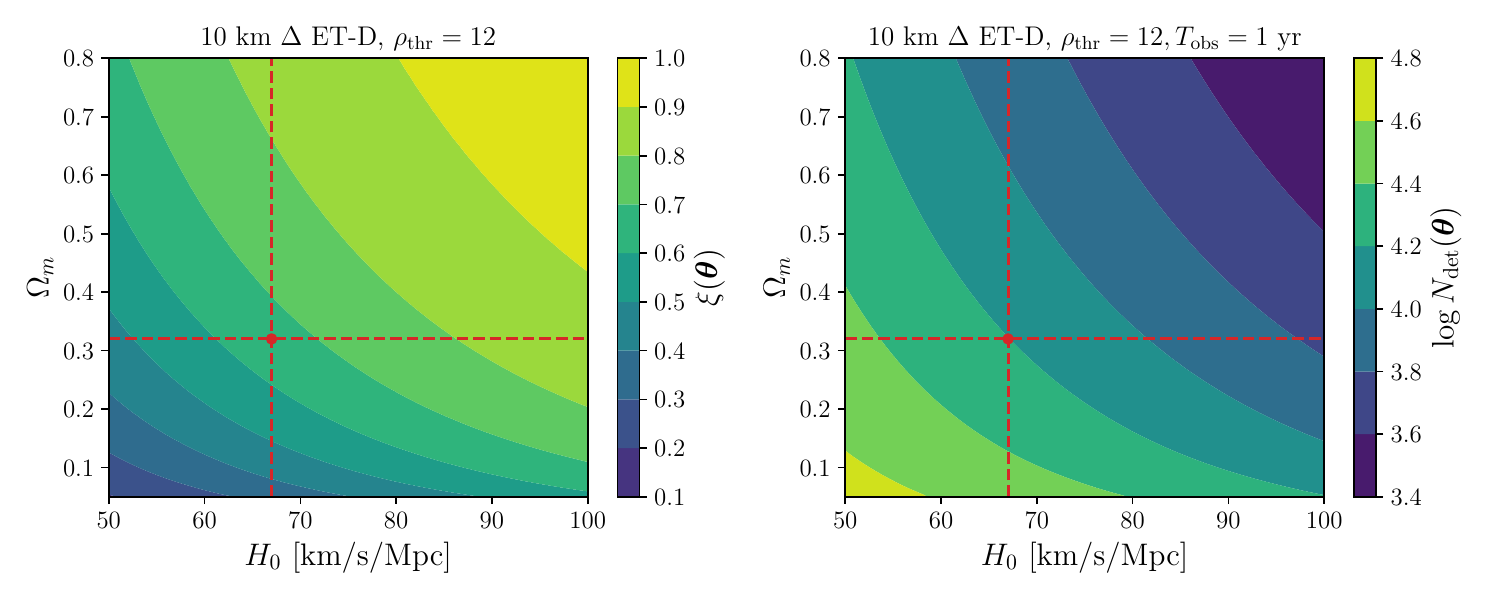}
    \caption{Fraction and number of detectable events for the triangular ET design for a large range of cosmological parameters fixing the fiducial astrophysical parameters (\autoref{tab:fiducial}); while red dashed lines represent fiducial cosmological parameters.}
    \label{fig:xi}
\end{figure}

In \autoref{fig:xi} we show the fraction of events varying cosmological parameters on a fine grid for ET reference sensitivity.

\subsection{Computing the likelihood}

Finally, to complete the hierarchical likelihood of \autoref{eq:final_hier_like} we want to implement the following expressions
\begin{equation}
    \int \text{d}z\,p(d_L^{i}|d_L(z|\thet))\,P(z|\thet)\,.
\end{equation}
The expression for $p(d_L^{i}|d_L(z|\thet))$ cannot be arbitrary because $p(\text{det}|d_L(z,\thet))$ constraints this expression according to \autoref{eq:p_det}, i.e.
\begin{equation}
    p({\rm det}|d_L(z,\thet)) = \int_{\tilde{d}_L\in \text{detectable}} \text{d}\tilde{d}_L\,p(\tilde{d}_L|d_L(z,\thet))\,,
\end{equation}
with $\tilde{d}_L$ the measured luminosity distance; moreover, the left hand side of the equation enters into $\xi(\thet)$ as expressed in \autoref{eq:csi_def}, therefore to avoid a biased posterior it is necessary to ensure consistency between these terms.

In a mock analysis the expression for $p(d_L^{i}|d_L)$ is set by the function with which the dataset is created, and in this case we have a normal distribution centered on the true value\footnote{In a real data scenario the gravitational-wave likelihood is not analytical. It must instead be reconstructed from posterior samples and inference priors, and a resampling method is then used to span the entire population-parameter prior space \cite{LIGOScientific:2026uyd,LIGOScientific:2026ctl}.} (see point 6. from \autoref{sec:datasim}):
\begin{equation}
    p(\tilde{d}_L|d_L(z,\thet),\sigma_i) = \mathcal{N}(\tilde{d}_L|\mu = d_L(z,\thet), \sigma = \sigma_i)\,.
\end{equation}
However, this expression is still inconsistent for two reasons:
\begin{itemize}
    \item[1)] The expression for $\sigma_i$ contains information on the true luminosity distance, which also enters in the expression of the SNR and it should be free to vary similarly to the mean $\mu$ of the extraction function. As noticed in \cite{Gair:2022zsa}, using a fixed luminosity distance in the scatter can lead to a bias in the posterior, therefore we generalize the scatter as
    \begin{equation}
        \sigma(z,\thet, \tilde{\omeg}_0) = \dfrac{2 d_L(z,\thet)}{\rho(z,\thet,\tilde{\omega}_0)}\,,
    \end{equation}
    where we also explicitly show the dependence on other parameters like masses and angles that are fixed in this context.
    \item[2)] We can expect that particular combinations of $(z,\thet)$ determine an SNR below detection threshold and accepting these points is in contrast with our SNR criterion. The solution is to impose a truncation whenever $\rho(z,\thet,\tilde{\omeg}_0)<\rho_{\rm thr}$ using a step function $\Theta(\rho(z,\thet,\tilde{\omega}_0) - \rho_{\rm thr})$.
\end{itemize}
Finally, the unbiased expression for $p(d_L^i|d_L)$ is thus:
\begin{equation}
    p(d_L^i|d_L(z,\thet),\tilde{\omeg}_0) = \mathcal{N}(\tilde{d}_L|\mu = d_L(z,\thet), \sigma = \sigma(z,\thet, \tilde{\omeg}_0))\,\Theta(\rho(z,\thet,\tilde{\omega}_0) - \rho_{\rm thr})\,.
\end{equation}

Following \autoref{sec:datasim}, we create a mock dataset assuming an observation time of $T_{\rm obs}=1\,\rm{yr}$ and a SNR threshold of $\rho_{\rm thr}=12$.\footnote{For the triangular ET configuration, we compute the network SNR as $\rho^2 = \sum_{i=0}^{3}\rho_i^2$, with a corresponding threshold for each interferometer as $\rho_i\approx7$.} For the cosmological inference we consider a flat $\Lambda$CDM cosmology and $\thet = (H_0, \Omega_m)$, with flat priors $H_0 \in [60,\,76]~\rm{km}/\rm{s}/\rm{Mpc}$, $\Omega_m \in [0.2,\,0.45]$, and fix the other astrophysical parameters to the values in \autoref{tab:fiducial}. With this setup we computed the log likelihood of \autoref{eq:final_hier_like} on a fine grid of $200\times200$ to overcome the computational costs of a complete Monte Carlo Markov Chain (MCMC) sampling.

In \autoref{fig:hier_like} we present the posteriors on the cosmological parameters using the full hierarchical likelihood $\mathcal{L}_F$ in \autoref{eq:final_hier_like} and we also show the contribution from number counts in \autoref{eq:number_like}, renamed $\mathcal{L}_N$ .

\section{The SBI approach: Marginal Neural Ratio Estimation}\label{sec:SBI}

SBI allows to perform inference without needing to explicitly evaluate the likelihood $\mathcal{L}(\{\xx\}|\thet)$ from \autoref{eq:hier_like}; instead, the likelihood is accessed implicitly via a
stochastic simulator that maps model parameters $\thet$ to data realizations $\xx$ \cite{Cranmer:2019eaq}. This is particularly advantageous in settings where the likelihood is computationally expensive or hard to express analytically, such as the case of interest here (\ie population inference of GW events in the
presence of selection effects). By
drawing parameters from the prior $p(\thet)$ and calling the simulator, one
generates $N$ data-parameter pairs $\{(\xx^1,\thet^1),\ldots,(\xx^N,\thet^N)\}$ drawn from the joint distribution $p(\xx,\thet)= p(\xx|\thet)p(\thet)$. These samples are subsequently used to train a neural network that directly estimates the posterior, the likelihood or the posterior-to-prior ratio \cite{Lueckmannetal21}. \

For the SBI analysis presented here, we developed the public code \Sireeni\footnote{\href{https://github.com/GFAbellan/Sireeni.git}{\texttt{https://github.com/GFAbellan/Sireeni.git}}}, which couples the \darksirens\ code described in \autoref{sec:data_ET}, used as the forward simulator, with the Marginal Neural Ratio Estimation (MNRE)  algorithm \cite{Miller:2021hys}, as implemented in the public code \Swyft\footnote{\href{https://github.com/undark-lab/swyft}{\texttt{https://github.com/undark-lab/swyft}}}. MNRE uses simulated data-parameter pairs to
train neural classifiers that directly learn 1D and 2D \emph{marginal posterior-to-prior ratios}. This allows to target only the marginal posteriors for the parameters of interest, while nuisance parameters are integrated out efficiently\footnote{In SBI, this marginalization is done implicitly by varying all the nuisance parameters during the simulation phase, but not passing them to the networks during training.} (see e.g. \cite{Cole:2021gwr, Montel:2022fhv, Saxena:2023tue,  Alvey:2023pkx,Bhardwaj:2023xph,FrancoAbellan:2024tbj, FrancoAbellan:2025fkb} for astrophysical or cosmological applications). We denote this marginal posterior-to-prior ratio as $r(\xx;\thet)$, with $\thet$ either a single or a pair of parameters from the set of parameters in \autoref{sec:analytical}. Using the Bayes' theorem,  we can write this quantity as
\begin{equation}
    r(\xx;\thet) \equiv \frac{p(\thet|\xx)}{p(\thet)} =  \frac{p(\xx|\thet)}{p(\xx)}  =  \frac{p(\xx,\thet)}{p(\xx)\,p(\thet)}.
    \label{eq:ratio_r}
\end{equation}
Determining
$r(\xx;\thet)$ is therefore equivalent to determining the
likelihood-to-evidence ratio, or the ratio of  densities for jointly drawn data-parameter pairs, $\xx,\thet \sim p(\xx,\thet)$, and marginally drawn pairs, $\xx,\thet \sim p(\xx)\,p(\thet)$. To estimate $r(\xx;\thet)$, we exploit the fact that, alongside the jointly drawn pairs, we can construct samples from the product of marginals  $p(\xx)\,p(\thet)$ by randomly shuffling the pair components. A binary classifier $d_\phi(\xx,\thet)$ (with $\phi$ denoting the learnable parameters of the neural network) is then trained to distinguish between two different classes: jointly drawn ($C=1$) and marginally drawn ($C=0$) data-parameter pairs,  \ie it is optimised to approximate
\begin{equation}
    d_\phi(\xx,\thet) \simeq p (C=1|\xx,\thet) = \frac{p(\xx,\thet)}{ p(\xx)p(\thet)+ p(\xx,\thet)}.
\end{equation}
This learning problem corresponds to minimising the binary cross-entropy loss
\begin{equation}
    \ell[d_\phi(\xx,\thet)] = -\int d\xx\, d\thet
    \left[ p(\xx,\thet)\ln d_\phi(\xx,\thet)
    + p(\xx)\,p(\thet)\ln\!\left(1-d_\phi(\xx,\thet)\right) \right],
    \label{eq:loss_function}
\end{equation}
whose functional minimisation yields the optimal classifier
$d^\star_\phi(\xx,\thet) = \sigma(\ln r(\xx;\thet))$, where $\sigma(x) = [1+\exp(-x)]^{-1}$ is the sigmoid function. In practice, the optimal network parameters $\phi$ are found by minimising a sample-based approximation of \autoref{eq:loss_function} using stochastic gradient descent. Then, the marginal posterior samples can be obtained by  drawing from the prior $p(\thet)$ and weighting by the trained ratios $r_\phi(\xx;\thet)$. \

Another important aspect of SBI is data compression. When working with high-dimensional data $\xx$ (as in this case, where GW catalogues contain $N_{\rm obs}\sim 10^4$ events), it is often advantageous to compress the data before the inference step. In the context of MNRE, the ratio estimation is therefore performed for $r (\bs, \thet)$, where $\bs = C_{\phi} (\xx)$ denotes a low-dimensional feature vector obtained via some compression network $C_{\phi}$. Both compression and inference networks are trained jointly by minimising the loss function in \autoref{eq:loss_function}. It can be shown that in this case the data summaries $\bs$ are learned such that they maximize the Jensen--Shannon divergence between the priors and the posteriors \cite{Cole:2021gwr}.

We will perform two types of SBI analyses: a \emph{cosmology-only} analysis that infers $(H_0, \Omega_m)$ with all astrophysical parameters fixed at their fiducial values, and a \emph{cosmology-plus-astrophysics} analysis that simultaneously infers $(H_0, \Omega_m, z_m, a, b)$. 

\subsection{Simulator and data compression}

Beyond the SBI algorithm itself, the implementation requires the specification of both a simulator and a data compression scheme tailored to the problem under consideration.\

\begin{figure}[h!]
    \centering
    \includegraphics[width=0.47\linewidth]{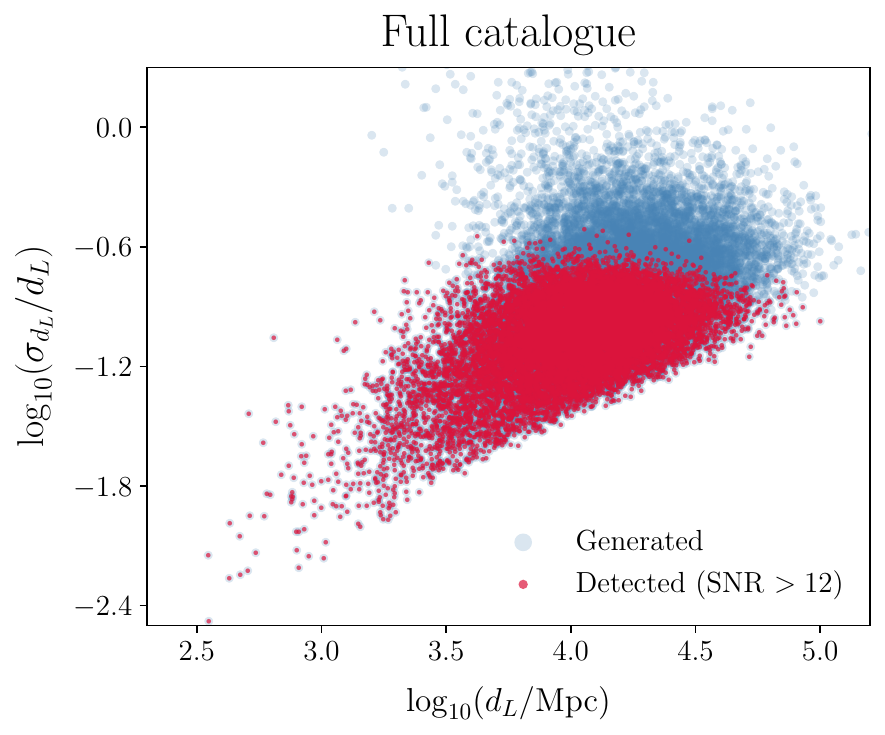}
    \hspace{5mm}
    \includegraphics[width=0.47\linewidth]{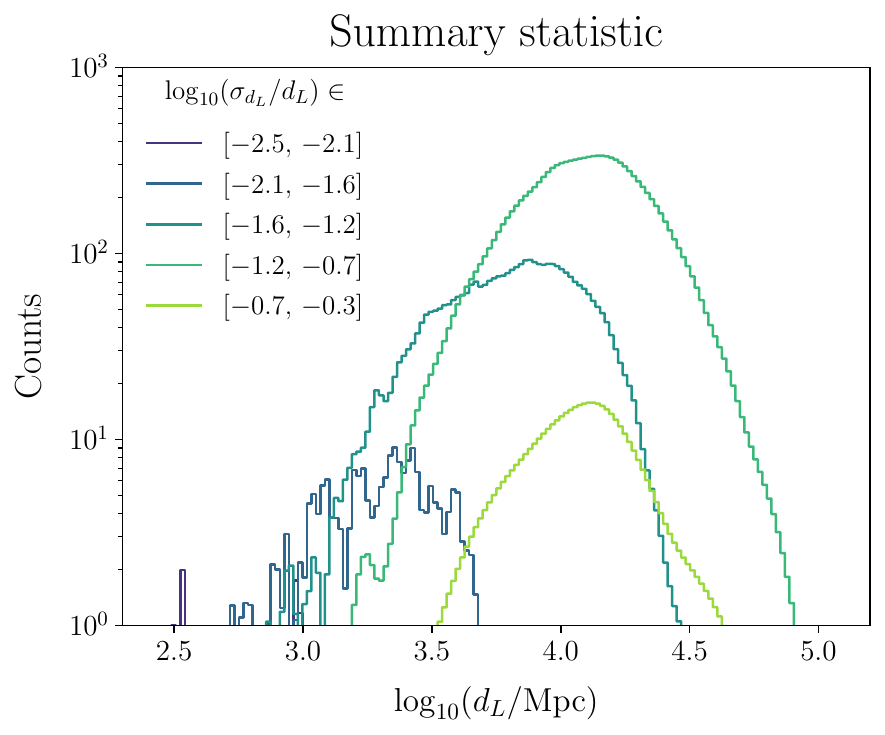}

    \caption{Illustration of the data compression scheme at the fiducial parameters of \autoref{tab:fiducial}.  \textit{Left:} Synthetic GW catalogue produced by \darksirens, showing the luminosity distance $d_L$ and fractional distance uncertainty $\sigma_{d_L}/d_L$ for each event. We show both the generated events and those that pass the $\mathrm{SNR} = 12$ detection threshold. \textit{Right:} The 2D soft histogram used as the main summary statistic in our SBI analyses. The detected events are hard-binned in $\log_{10}(\sigma_{d_L}/d_L)$ into five slices, and within each slice soft-binned into $200$ $\log_{10}(d_L)$ bins using a Gaussian kernel of width $\sigma_{d_L}$. 
    }
    \label{fig:soft_histogram}
\end{figure}

For the cosmology-only analyses, we sample the cosmological parameters
$\thet = (H_0, \Omega_m)$ from flat priors $H_0 \in [60,\,76]~\rm{km}/\rm{s}/\rm{Mpc}$ and $\Omega_m \in [0.2,\,0.45]$, and run the \darksirens\ code with all astrophysical parameters fixed to the fiducial values given in \autoref{tab:fiducial}.
For the cosmology-plus-astrophysics analysis we additionally vary the parameters $(z_m, a, b)$ over the flat priors $z_m \in [1.4,\,2.6]$,
$a \in [1.66,\,3.08]$, and $b \in [1.26,\,2.34]$ (corresponding to $\pm 30\%$ variations around their fiducial values)\footnote{These priors include a small region of the parameter space where $a < b$; in this regime the SFR density in \autoref{eq:SFR} can acquire negative (and hence unphysical) values at very low redshift. To avoid this, we simply enforce $N_{\rm obs} = 0$ for such parameter draws.}, while the amplitude $\nu$ remains fixed at its fiducial value since its effect on the inference is negligible (see \autoref{sec:datasim}). In all analyses we adopt $\rho_{\rm thr}=12$ and $T_{\rm obs}=1\,\rm{yr}$, matching the setup used in the HBI analysis.

Each simulation produces a catalogue with luminosity distances and their associated uncertainties
    \begin{equation}
        \xx = \left\{ d^i_L, \sigma^i_{d_L}\right\}_{i=1,...,N_{\rm obs}},
    \end{equation}
where $N_{\rm obs}$ is the number of events surviving the SNR cut. We generate two independent sets of $10^5$ simulations\footnote{We verified that a smaller simulation budget of $5\times10^4$ is already sufficient to obtain good results in the cosmology-only case; we nonetheless adopt $10^5$ simulations throughout as a conservative choice.}: one with all astrophysical parameters fixed at their fiducial values, used to train the cosmology-only networks, and one with $(z_m, a, b)$ also varied according to the priors above, used for the
cosmology-plus-astrophysics analysis.\

For data compression we extract two summary statistics from the simulated catalogues:
\begin{enumerate}
    \item The total number of observed events $N_{\rm obs}$.
    \item A 2D soft histogram $\bm{h}$ of shape $(N_{\rm frac}, N_{\rm bin}) = (5, 200)$.
\end{enumerate}
We refer to networks that jointly use both statistics as \emph{Number\,+\,Shape} networks,
reflecting their use of the total event count ($N_{\rm obs}$, the ``Number'' part) and the
distributional shape of the catalogue encoded in $\bm{h}$ (the ``Shape'' part).\

To construct the 2D soft histogram, events are first \emph{hard}-binned in $\log_{10}(\sigma_{d_L}/d_L)$ into $N_{\rm frac}=5$ equally-spaced slices covering the range $[-2.5,\,-0.3]$, with events outside this range assigned to the nearest edge bin. Hard-binning here means that each event is assigned to exactly one slice based on its observed value, with no uncertainty smearing.
Within each fractional-uncertainty slice, events are \emph{soft}-binned in
$\log_{10}(d_L/\mathrm{Mpc})$ into $N_{\rm bin}=200$ bins uniformly spaced over
$[2.0,\,5.5]$.\footnote{We note that $N_{\rm frac}=5$ and $N_{\rm bin}=200$ were chosen by trial-and-error: we found that coarser and moderately finer binnings both gave somewhat worse performance.} Unlike hard-binning, soft-binning spreads each event across neighbouring bins with weights proportional to the probability of its true distance lying within each bin.

The weight of event $i$ in $\log_{10}(d_L)$ bin $j$, i.e.\ the probability for event $i$ to lie within that bin given its measured distance and uncertainty, is \begin{equation}
w_{ij} = \int_{d_L^{\rm low,j}}^{d_L^{\rm high,j}} \mathcal{N}\!\left(d_L \,\middle|\, \mu =d_L^i, \sigma = \sigma^i_{d_L}\right) \, \mathrm{d}d_L
        = \Phi\!\left(\frac{d_L^{\rm high,j} - d_L^i}{\sigma^i_{d_L}}\right) - \Phi\!\left(\frac{d_L^{\rm low,j} - d_L^i}{\sigma^i_{d_L}}\right),
    \end{equation}
where $\Phi$ is the standard normal cumulative distribution function, and $[d_j^{\rm low},d_j^{\rm high}]$ denote the edges of each bin $j$. With this, the count in each cell of the 2D histogram is the sum of weights over all events assigned to
the corresponding fractional-uncertainty slice. The 2D histogram retains the per-event correlation between $d_L$ and $\sigma_{d_L}/d_L$, which a standard 1D $d_L$ histogram would discard.\footnote{This correlation proves essential for constraining the cosmological parameters in the cosmology-plus-astrophysics case: we verified that a 1D soft histogram in $\log_{10}(d_L)$ with the same 200 bins performs comparably to the 2D version for the cosmology-only analyses, but fails to provide sufficient constraining power on  $(H_0,\Omega_m)$ when the astrophysical parameters $(z_m, a, b)$ are simultaneously inferred, as their imprint on the GW catalogue is partly encoded in the $\sigma_{d_L}/d_L$ distribution.} \autoref{fig:soft_histogram} illustrates the summary statistic at the fiducial parameters:
the left panel shows the raw synthetic GW catalogue in the
$(\log_{10}(d_L),\,\log_{10}(\sigma_{d_L}/d_L))$ plane, and the right panel shows the
resulting 2D soft histogram.

The resulting $(5\times 200)$-dimensional array is flattened and concatenated with $N_{\rm obs}$
to form the full $1001$-dimensional summary vector $\tilde{\bs} = (N_{\rm obs},\,\bm{h})$. The summary vector $\tilde{\bs}$ is first passed through an online normalisation layer and
then compressed by a multi-layer perceptron (MLP) whose architecture is tuned separately for each analysis via hyperparameter optimisation with \texttt{Optuna} \cite{Akiba:2019lwq}. The compression MLP for the Number\,+\,Shape analysis has one hidden layer with 384 neurons and produces 2 compressed features per cosmological parameter in the cosmology-only setup; when astrophysical parameters are also varied, it instead has three hidden layers with 256, 228, and 203 neurons, producing 3 compressed features per inferred parameter. In both cases the marginal ratio estimators for the 1D and 2D posteriors are implemented as MLPs with four residual blocks of 64 neurons each.

We additionally train a \emph{Number}-only network  (using $N_{\rm obs}$ as its sole input) for the cosmology-only case, allowing a direct comparison with the HBI results obtained from the number-count term $\mathcal{L}_N$. Its compression MLP has a single hidden layer with 64 neurons and maps the normalised
$N_{\rm obs}$ to one compressed feature per parameter, which is then fed separately to the ratio estimators. All other network settings are identical to those above.

\subsection{Training strategy}

All networks are trained with the \texttt{AdamW} optimiser together with a
\texttt{ReduceLROnPlateau} scheduler that reduces the learning rate (lr) by a factor of 10
whenever the validation loss does not improve over 3 consecutive epochs.
Training is stopped early if the validation loss does not improve over 5 epochs, and the
checkpoint with the lowest validation loss is retained.
We use a batch size of 64 and reserve 20\% of the simulated samples for validation.
The initial learning rate and weight decay ($\lambda$) are tuned independently for each network via
\texttt{Optuna}: for the cosmology-only Number\,+\,Shape network the tuned values are
$\mathrm{lr}=6.8\times10^{-5}$ and $\lambda=0$;
for the Number-only network, $\mathrm{lr}=1.2\times10^{-4}$ and $\lambda=0$;
and for the varying astrophysics Number\,+\,Shape network,
$\mathrm{lr}=2.3\times10^{-4}$ and $\lambda=2.5\times10^{-6}$.

\section{Results}\label{sec:results}

\begin{figure}[h!]
    \centering
    \includegraphics[width=0.6\linewidth]{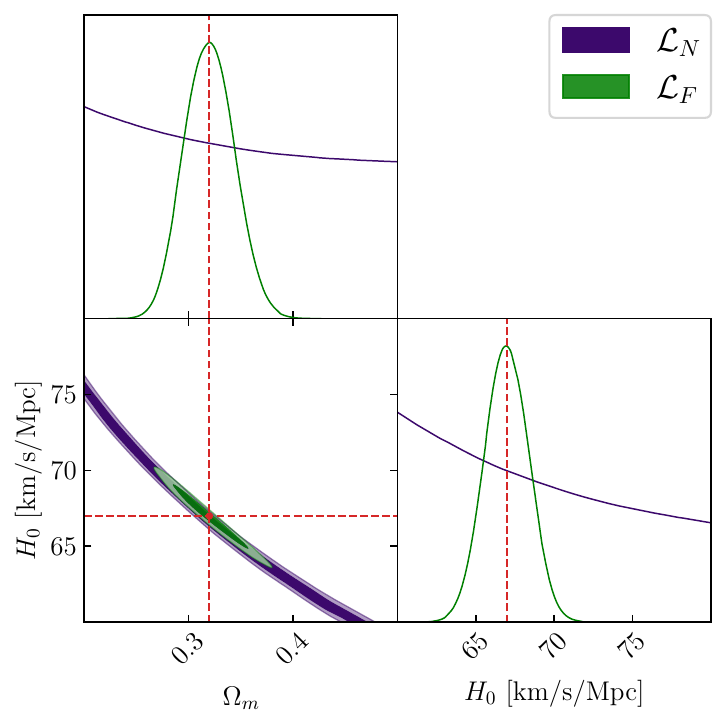}
    \caption{Cosmological posteriors from one year of observations of ET 10 km triangular configuration. In \emph{purple} we have the Poisson contribution to the hierarchical likelihood, namely $\mathcal{L}_N$ (\autoref{eq:number_like}), and in \emph{green} the complete hierarchical likelihood $\mathcal{L}_F$ (\autoref{eq:final_hier_like}). The red dashed lines represent the injected fiducial values from \autoref{tab:fiducial}}
    \label{fig:hier_like}
\end{figure}

We first validate the traditional hierarchical approach on its own. In \autoref{fig:hier_like} we show the posteriors on $(H_0,\Omega_m)$ obtained from the full hierarchical likelihood $\mathcal{L}_F$ (\autoref{eq:final_hier_like}), compared to the contribution given by the number-counts only term $\mathcal{L}_N$ (\autoref{eq:number_like}). As already shown in Ref. \cite{Antinozzi:2023yvl}, $\mathcal{L}_N$ alone constrains only a broad, highly degenerate strip in the $(H_0,\Omega_m)$ plane, since the total number of detected events depends on the cosmological parameters mainly through the comoving volume and cannot break the degeneracy between these two parameters. Including the rest of the likelihood with information carried by the individual luminosity-distance measurements collapses this degeneracy substantially, yielding a well-constrained 2D posterior centred on the fiducial values of \autoref{tab:fiducial}. When the full likelihood is considered, we obtain $H_0=67.0\pm1.3$ km/s/Mpc and $\Omega_m=0.32^{+0.021}_{-0.025}$. While not competitive with results such as those from DESI+BBN \cite{DESI:2025zgx}, which yields errors almost five times smaller, this result highlights the relevance of the Einstein Telescope capability, as it will provide a completely independent estimate of cosmological parameters.

\begin{figure}[h!]
    \centering
    \includegraphics[width=0.45\linewidth]{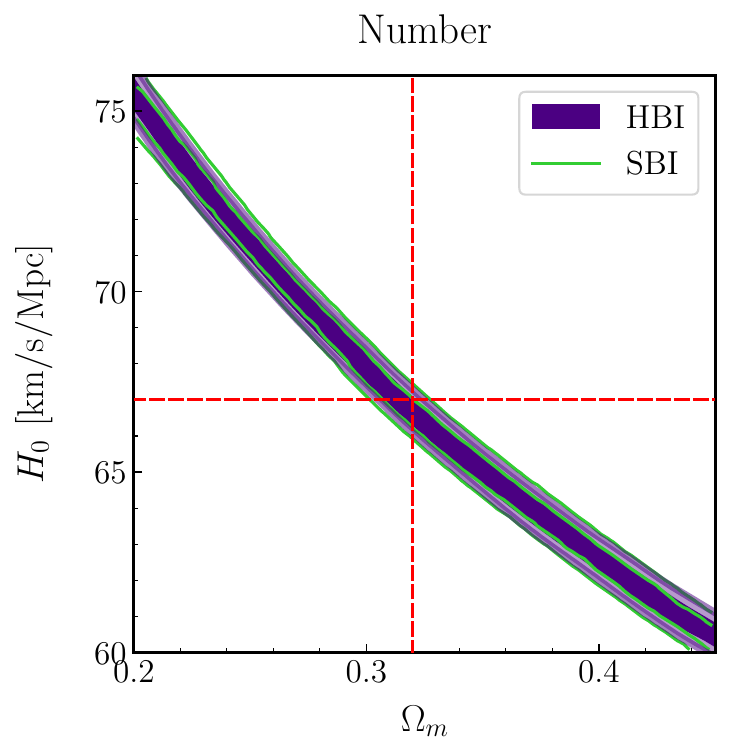}
    \hspace{10mm}
    \includegraphics[width=0.44\linewidth]{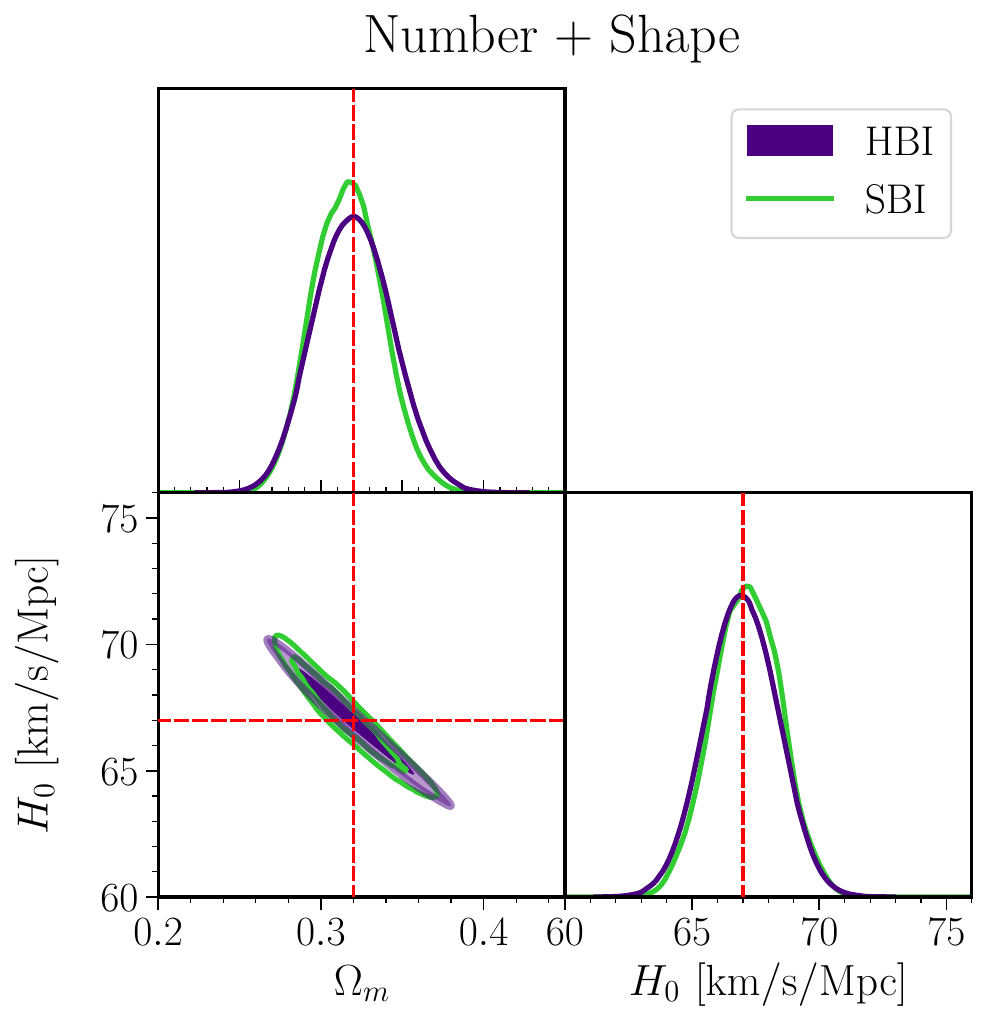}
    \caption{Cosmological posteriors from one year of observation of ET, using both an explicit hierarchical likelihood (purple contours) and SBI with our \Sireeni~code 
    (green contours). The left panel uses only the observed number of GW events, while the right panel additionally incorporates information from the distribution of events in luminosity-distance.  The red dashed lines indicate the injected values. In both cases, the SBI results show excellent agreement with the HBI approach.}
    \label{fig:swyft_vs_mcmc}
\end{figure}

We then compare the SBI pipeline, run with our public code \Sireeni, against this traditional benchmark. \autoref{fig:swyft_vs_mcmc} shows the posteriors obtained with MNRE (green) overlaid on the analytical HBI posteriors (purple), for two choices of summary statistic: using only the number of observed events $N_{\rm obs}$ (left panel), and using the full Number+Shape summary, i.e. $N_{\rm obs}$ together with the 2D soft histogram of $(d_L,\sigma_{d_L}/d_L)$ (right panel). In both cases the SBI posteriors reproduce the analytical contours essentially one-to-one, with the injected fiducial values (red dashed lines) recovered consistently by both methods. This agreement confirms that MNRE, trained on $10^5$ forward simulations, is able to extract the same cosmological information as the exact hierarchical likelihood without requiring an explicit likelihood evaluation, and that our two-dimensional summary statistic retains the relevant information originally encoded in the per-event $(d_L,\sigma_{d_L})$ pairs.

\begin{figure}[h!]
    \centering
    \includegraphics[width=0.9\linewidth]{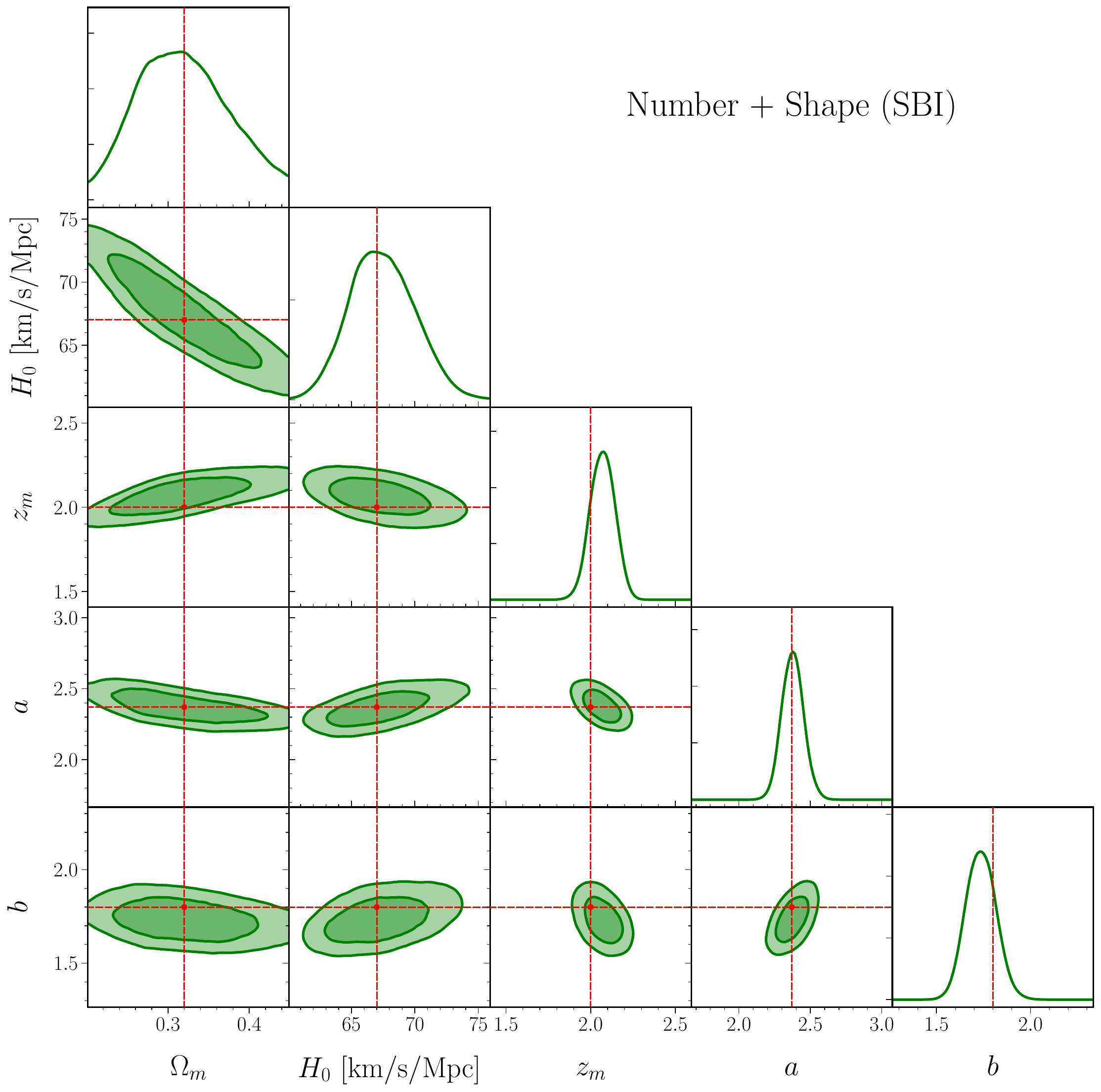}
    \caption{ Cosmological and astrophysical posteriors from one year of observation of ET, using  SBI with our \Sireeni~code. We incorporate information from the observed number of GW events as well as the distribution of events in luminosity-distance. The red dashed lines indicate the injected values. }
    \label{fig:swyft_cosmo+astro}
\end{figure}

Having validated the method against the tractable cosmology-only case, we exploit the ability of SBI to incorporate additional parameters into the forward simulator at negligible extra analytical cost, to extend the inference to the astrophysical parameters $(z_m, a, b)$ that shape the redshift distribution through Eq.~(\ref{eq:SFR}). \autoref{fig:swyft_cosmo+astro} shows the resulting five-parameter posterior for the analysis with varying astrophysics, obtained from the Number+Shape summary statistic.  The joint posterior
recovers all five injected values within the plotted credible regions. We stress that extending the analysis in this way required only minor changes to the SBI pipeline: sampling the enlarged prior volume and a small adjustment of the compression-network architecture, but no change to the simulator or to the inference algorithm itself. This is in marked contrast to the HBI framework, where the same extension would demand the additional importance-sampling dimensions and the more general SNR-scaling treatment discussed in \autoref{sec:xi}.  Regarding the scientific
conclusion, once the merger-rate parameters are marginalised over, the constraints on $(H_0, \Omega_m)$ degrade appreciably with respect to the fixed-astrophysics case. This degradation of constraining power reflects the strong degeneracies between the cosmological and astrophysical parameters, highlighting the importance of inferring them jointly.

In addition, we performed empirical coverage tests to validate the statistical behaviour of the estimated posteriors for all the SBI analyses presented above. As detailed in \appref{sec:coverage_test}, we find that the trained ratio estimators are well calibrated in every case.

Beyond the statistical agreement, the two approaches differ markedly in computational cost. Evaluating the full hierarchical likelihood on a $200\times200$ grid in $(H_0,\Omega_m)$ requires roughly $370$ CPU-hours, dominated by the precomputation of the theoretical quantities ($P(z|\thet)$, $d_L(z|\thet)$ and the SNR) on the grid, together with about $10$ GPU-hours for the injection campaign needed to estimate the selection function. The SBI pipeline, in the same cosmology-only setup, has a cost of roughly $175$ CPU-hours to generate the forward simulations, and additionally requires about $20$ GPU-minutes to train the networks (we provide a detailed breakdown of these estimates for both methods in \appref{sec:time_performance}). The two upfront costs are therefore of similar order, but they scale very differently. For the HBI method, any extension to additional population parameters inflates the cost significantly through the higher-dimensional resampling it demands, whereas the SBI cost is essentially insensitive to such extensions. Moreover, SBI is fully amortized: once trained, the networks yield posteriors for new mock catalogues in seconds, with no further simulation. Together, these properties make SBI considerably more scalable to the higher-dimensional population models and larger event catalogues expected in the ET era.

\section{Conclusions}\label{sec:conclusions}

In this work we compared two approaches to GW population inference applied to dark sirens cosmology in view of the Einstein Telescope: the traditional hierarchical Bayesian likelihood and simulation-based inference via Marginal Neural Ratio Estimation.\

Using a simplified but realistic mock ET catalogue of $O(10^4)$ binary black hole events generated with \texttt{darksirens}, we showed, using our public code \Sireeni, that SBI reproduces the cosmological posteriors obtained from the exact hierarchical likelihood with excellent accuracy, both when using only the number of detected events and when including the full information of the luminosity-distance distribution. We further used empirical coverage tests to check the statistical consistency of the trained networks.\

Beyond validating the equivalence of the two methods in a tractable regime, we highlighted their complementary strengths. The hierarchical likelihood remains exact and interpretable, but its evaluation at this stage is dominated by a computationally expensive injection campaign whose cost grows quickly with the number of parameters through the importance-sampling and resampling procedure required to estimate the selection function and at a later stage can be additionally burdened by the computation of each event-integral in Eq.~(\ref{eq:final_hier_like}). SBI trades this analytical control for a forward-simulation approach that is comparatively inexpensive to extend: incorporating the astrophysical merger-rate parameters $(z_m,a,b)$ alongside $(H_0,\Omega_m)$ required no conceptual change to the pipeline, and the resulting five-parameter posterior recovered all injected values consistently. This flexibility, together with the amortized nature of the trained networks which, once trained, can be reused for repeated or extended inference without recomputing the likelihood, makes SBI a promising tool for the higher-dimensional population analyses and larger event catalogues that third-generation detectors like ET will provide.\

Several simplifications adopted here point to natural extensions of this work. We fixed the black hole component masses to a monochromatic value in order to isolate the redshift/luminosity-distance sector of the inference; relaxing this assumption to a full mass spectrum and mass-ratio distribution, while only mildly complicating the SBI simulator, would substantially increase the dimensionality of the importance sampling required by the hierarchical approach, and would offer a more stringent test of the relative scalability of the two methods. Similarly, extending the comparison to alternative ET geometries, would allow this framework to be tested in more general scenarios. Finally, applying the same pipeline to real gravitational-wave catalogues, where the single-event likelihood is not analytical and must instead be reconstructed from posterior samples, will be an important next step in establishing SBI as a robust and scalable alternative to hierarchical Bayesian inference for spectral siren cosmology.

\section*{CRedIT statement}

\textbf{G. Antinozzi}: Conceptualization; Methodology; Software; Visualization; Writing – original draft; Writing – review \& editing;
\textbf{G. Franco Abell\'an}: Conceptualization; Methodology; Software; Visualization; Writing – original draft; Writing – review \& editing;
\textbf{D. Sciotti}:
Methodology; Software; Writing – review \& editing;
\textbf{M. Martinelli}: Conceptualization; Methodology; Software; Writing – original draft; Writing – review \& editing.

\acknowledgments

MM acknowledges funding by the Agenzia Spaziale Italiana (\textsc{asi}) under agreement n. 2024-10-HH.0 and support from INFN/Euclid Sezione di Roma. GA and MM thank Simone Mastrogiovanni for useful discussions on hierarchical inference. GA gratefully acknowledges SISSA for the Ph.D. scholarship and thanks M.~Spera and A.~Lapi for their insightful feedback as supervisors. GFA gratefully acknowledges the computer resources at Artemisa, funded by the European Union ERDF and Comunitat Valenciana as well as the technical support provided by the Instituto de Fisica Corpuscular, IFIC (CSIC-UV).

\appendix

\section{Coverage tests for SBI}
\label{sec:coverage_test}

A useful feature of many SBI methods is that they yield amortized inference networks: once trained, they can produce a posterior estimate not just for the real data (as MCMC does), but for any mock observation drawn from the prior. This property makes it possible to evaluate statistical diagnostics such as the empirical coverage \cite{Hermans:2021rqv},  i.e. how often a stated credible region actually captures the true parameter value. For a well-calibrated posterior, the $p\%$  credible interval contains the true value $p\%$  of the time, and the empirical coverage plotted versus the nominal level follows a straight diagonal. Diagnostics of this kind are typically out of reach for conventional likelihood-based approaches, which would demand hundreds of independent MCMC runs, one for every mock dataset entering the coverage plot.\

\begin{figure}[h!]
    \centering
    \includegraphics[width=0.45\linewidth]{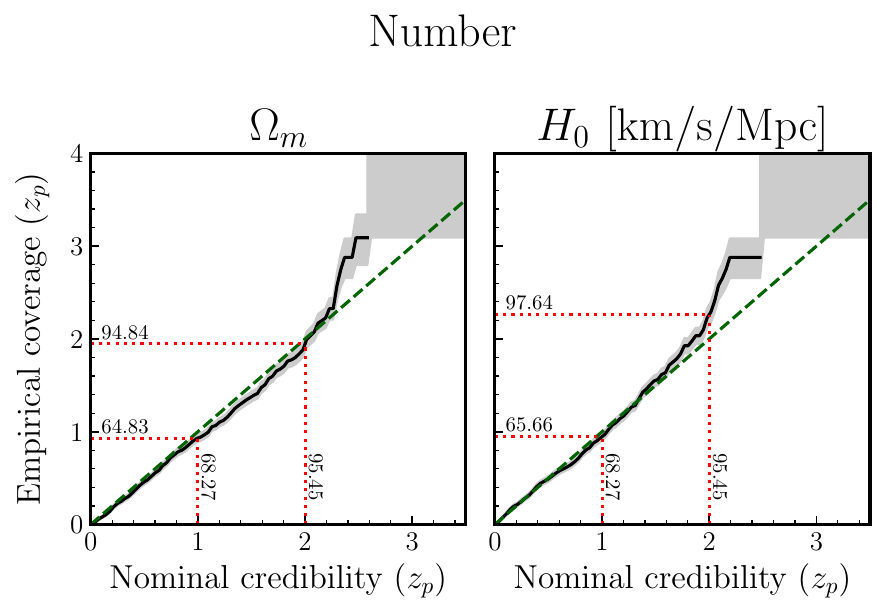}
    \hspace{5mm}
    \includegraphics[width=0.45\linewidth]{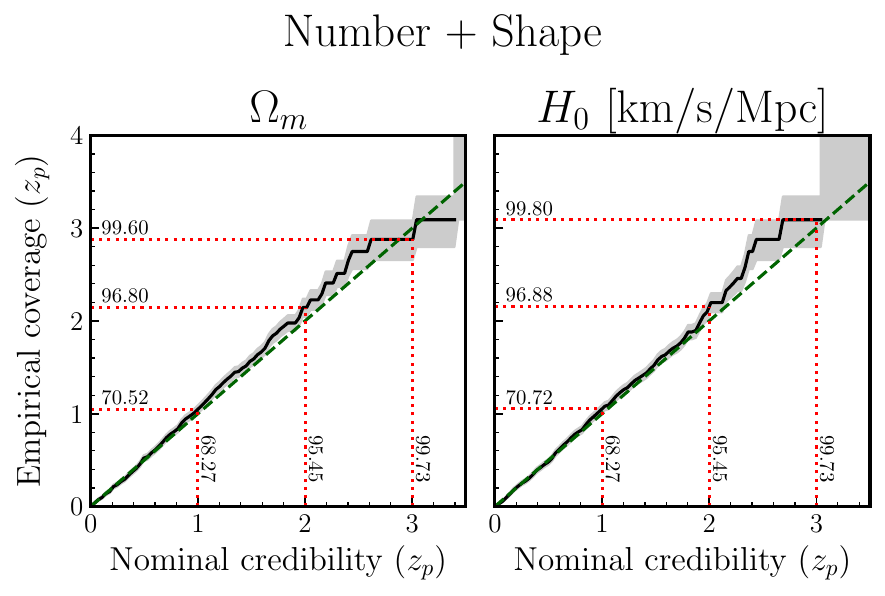}
    \caption{Coverage test for the cosmological parameters from the Number (left) and Number + Shape (right) analyses, assuming fixed astrophysical parameters.}
    \label{fig:coverage_test_swyft_1}
\end{figure}

\begin{figure}[h!]
    \centering
    \includegraphics[width=1.0\linewidth]{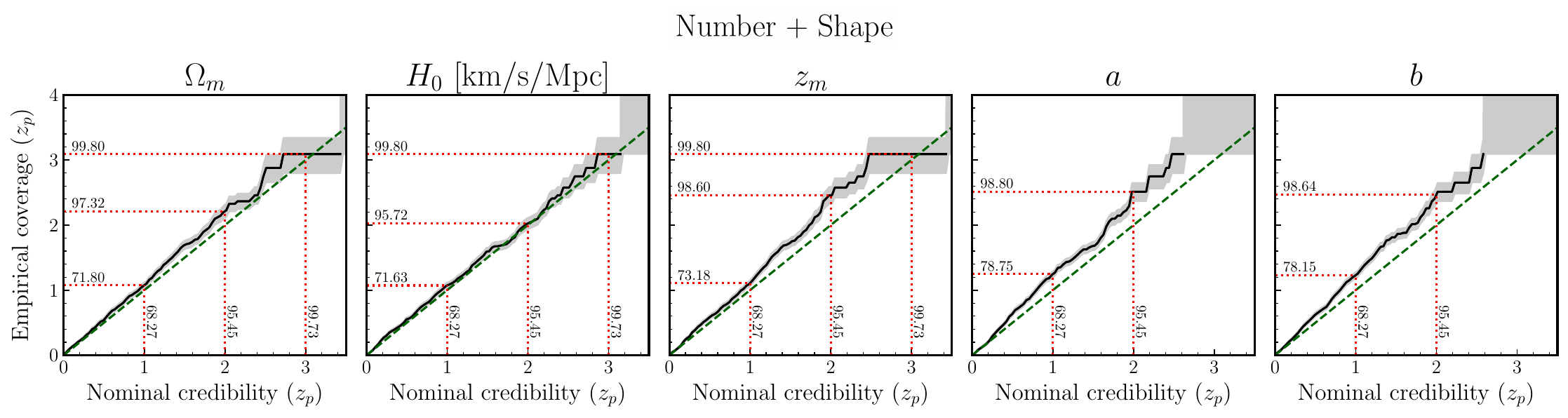}
    \caption{Coverage test for the  parameters from the Number + Shape analysis with three varying astrophysical parameters.}
    \label{fig:coverage_test_swyft_2}
\end{figure}

We perform a coverage test for the marginal 1D posteriors of the different SBI analyses shown in the main text, using a batch of $500$ simulations in each case. Instead of reporting the highest-posterior-density region $p$ itself, we work with a transformed variable $z_p$ given by $p/100 = \frac{1}{\sqrt{2\pi}}\int_{-z_p}^{z_p}dz \exp{(-z^2/2)}$, which places greater emphasis on the tails of the posteriors.  With this choice, the familiar  $(1, 2, 3)\sigma$  regions map onto $z_p = (1, 2, 3)$, corresponding to $p = (68.27, 95.45, 99.73)$. We additionally quantify the uncertainty on the empirical coverage due to the finite sample size via the Jeffreys interval \cite{Cole:2021gwr}. The results of the coverage test are shown in \autoref{fig:coverage_test_swyft_1} for the Number and Number+Shape analyses in the cosmology-only setup, and in \autoref{fig:coverage_test_swyft_2} for the Number+Shape analysis when astrophysical parameters are also varied. We find close agreement between the empirical coverage and the expected confidence levels across all parameters. This indicates that the trained ratio estimators produce well-calibrated posteriors, supporting the validity of the SBI constraints presented in the main text.

\section{Time performance: SBI vs. likelihood}
\label{sec:time_performance}

In this appendix we detail and compare the computational cost of the two approaches. All timings refer to the setup used in the main analysis: a $200\times200$ grid in $(H_0,\Omega_m)$ for the hierarchical method, and $10^5$ forward simulations per training set for SBI.

\subsection*{Hierarchical Bayesian inference}

For the HBI method the generation of the injection campaign takes most of the time. In a naive approach without any resampling, the computation of the necessary number of injected events ($N_{\rm gen}=10^8$) for a single set of parameters takes $10^4$ s on a single CPU using \texttt{darksirens}. Quantities such as $P(z|\thet),d_L(z|\thet), \rho(d_L(z|\thet),\tilde{\omeg})$ can be precomputed and they take respectively $0.1$ s, $0.03$ s and $5.4$ s per parameter space point. If the precomputation of $\xi,\ P(z|\thet),\ d_L(z|\thet),\ \rho(d_L(z|\thet),\ \tilde{\omeg})$ is achieved (practically dominated by the injection time), then the evaluation of $\mathcal{L}_N$ on a single parameter point takes $5$ ms, whereas it takes $1$ s for $\mathcal{L}_S$, both on a single CPU. The total computation time sums up to roughly $3$ hours for a single likelihood evaluation.

To overcome this computational wall, we obtained the results using a grid likelihood in the cosmological parameter space and parallelly computed both the necessary theoretical functions and $\xi$, the latter also derived with a importance sampling (IS) and resampling approach as described in \autoref{sec:xi}. The computation times in this case were:
\begin{itemize}
    \item[i)] $12$ hours to compute $P(z|\thet)$, $d_L(z|\thet)$, $\rho(d_L(z|\thet),\tilde{\omeg})$
   for a  $(H_0,\Omega_m)$ grid, using 30 CPUs;

    \item[ii)] following \autoref{sec:xi} we have $94$ s to compute $10^7$ injections for a reference parameter $\tilde{\thet}$ using 30 CPUs, and $1$ hour of IS and resampling to compute $\xi$ on the full cosmological grid using \texttt{JAX} \cite{jax2018github} and one GPU ({\tt NVIDIA GeForce RTX 4070 mobile}), then repeated 10 times to obtain $10^8$ injections per grid point;
    \item[iii)]  with precomputed quantities we got a faster computation of $\mathcal{L}_S$ using \texttt{JAX} and a GPU at $8$ ms per cosmological space point.
\end{itemize}
Using a grid likelihood with CPU parallelization and a GPU acceleration, brings the total total wall-clock time for the entire grid to $\approx 22.5$ hours, thus a single likelihood evaluation of $\approx 2$ s per grid point. Accounting for the number of cores employed, the dominant term (i) corresponds to $360$ CPU-hours, to which the injection step (ii) adds $\simeq 8$ CPU-hours, for a total of $\approx370$ CPU-hours.  The GPU cost is $\approx10$ GPU-hours, dominated by the IS and resampling step.

\subsection*{Simulation-based inference}

For the SBI approach, generating a set of $10^5$ simulations with \darksirens\ took $\sim2.5$ h using 70 CPU cores in parallel, i.e. $\approx175$ CPU-hours. The cosmology-only analysis relies on a single such set, shared by both the Number-only and Number+Shape networks. Training the inference networks additionally required $\sim10$ min on a single GPU ({\tt NVIDIA A100-SXM4 40GB}) per network, i.e. $\approx20$ GPU-minutes for the two cosmology-only networks.\

The analysis with varying astrophysics requires a second, independent set of $10^5$ simulations (a further $\approx175$ CPU-hours) and one additional network ($\sim10$ GPU-minutes), but the HBI method was not extended to this case since it would have been computationally prohibitive. Crucially, thanks to the amortized nature of SBI, inference on a new dataset requires no further simulation or training and is performed in a matter of seconds.

\bibliographystyle{JHEP}
\bibliography{references.bib}

\end{document}